# Tabular Two-Dimensional Correlation Analysis for Multifaceted Characterization Data

Shun Muroga[1*], Satoshi Yamazaki[2], Koji Michishio[3], Hideaki Nakajima[1], Takahiro Morimoto[1], Nagayasu Oshima[3], Kazufumi Kobashi[1], Toshiya Okazaki[1]

1: Nano Carbon Device Research Center, National Institute of Advanced Industrial Science and Technology (AIST), Tsukuba Central 5, 1-1-1, Higashi, Tsukuba, Ibaraki 305-8565, Japan
2: Research Association of High-Throughput Design and Development for Advanced Functional Materials (ADMAT), Tsukuba, Ibaraki 305-8565, Japan
3: Research Institute for Measurement and Analytical Instrumentation, National Institute of Advanced Industrial Science and Technology (AIST), Tsukuba Central 2, 1-1-1, Umezono, Tsukuba, Ibaraki 305-8568, Japan
*Corresponding author: Shun Muroga, Nano Carbon Device Research Center, National Institute of Advanced Industrial Science and Technology (AIST), E-mail: muroga-sh@aist.go.jp

**Abstract**
We propose tabular two-dimensional correlation analysis for extracting features from multifaceted characterization data, essential for understanding material properties. This method visualizes similarities and phase lags in structural parameter changes through heatmaps, combining hierarchical clustering and asynchronous correlations. We applied the proposed method to datasets of carbon nanotube (CNTs) films annealed at various temperatures and revealed the complexity of their hierarchical structures, which include elements like voids, bundles, and amorphous carbon. Our analysis addresses the challenge of attempting to understand the sequence of structural changes, especially in multifaceted characterization data where 11 structural parameters derived from 8 characterization methods interact with complex behavior. The results show how phase lags (asynchronous changes from stimuli) and parameter similarities can illuminate the sequence of structural changes in materials, providing insights into phenomena like the removal of amorphous carbon and graphitization in annealed CNTs. This approach is beneficial even with limited data and holds promise for a wide range of material analyses, demonstrating its potential in elucidating complex material behaviors and properties.

## Introduction
The materials that surround us exhibit hierarchical structures, with structural changes at each scale impacting the bulk properties. Analyzing such hierarchical structures at different length-scales is crucial for revealing the phenomena occurring in materials. For example, a comprehensive understanding of polymeric materials, in principle, would require evaluations of structures spanning the chemical bonds to crystals, including primary repeating structures, molecular weight distribution, and lamellar structures formed by folded chains, and spherulites[1]. This multifaceted analysis and unraveling of phenomena in materials is a universal practice, not limited to a specific material. However, multifaceted characterization data often contain changes where multiple factors exert complex effects in different sequences, making it challenging to arrive at an clear understanding through simple analysis. Consequently, there is a growing demand for analytical methods capable of effectively extracting information from multifaceted characterization data.

To establish methods for analyzing multifaceted characterization data, various approaches using data science have been pursued. One major obstacle that complicates analysis is multi-collinearity, where multiple factors change simultaneously, making it difficult to recognize behaviors by through visual inspection of the simple plotting of data. A typical approach to address such multi-collinearity is regression analysis that can recognize this such behavior. Partial least squares (PLS) regression is aptly suited for this task[2]. PLS regression can be applied not only to commonly used vibrational spectroscopy data[3,4] but also to multifaceted analytical data with strong multi-collinearity[5,6]. For instance, applying PLS regression to a multifaceted dataset of carbon nanotube yarns has led to successful differentiation of contributions to mechanical strength and conductivity[5]. Additionally, recent advancements in deep learning and generative AI have enabled the development of multimodal deep learning, which integrates different types of multifaceted characterization data to predict material properties holistically[7]. These analytical methods are extremely effective in optimizing material properties, but challenges remain in understanding the sequence and impact of multiple factors affecting the multifaceted characterization data.

It is well known that two-dimensional correlation analysis is effective in evaluating the sequence of fluctuating factors from changes in analytical data. Especially due to its effectiveness in analyzing vibrational spectroscopy spectra and *in-situ* measured spectra, this method, also referred to as two-dimensional correlation spectroscopy (2DCOS), has been actively researched worldwide[8–18]. In 2DCOS, the extent and sequence of changes in absorption bands can be quantified by measuring the degree of change in sync (synchronous) or with a phase lag (asynchronous) in response to environmental or situational perturbations, such as temperature or strain. The use of synchronous and asynchronous spectra for feature extraction makes this method highly sensitive to detecting peak shifts or broadening due to interactions. In addition, its application has successful in analyzing phenomena, such as degradation[19–21], curing reaction[22], deformation of crystalline and amorphous regions[23,24], and interactions of proteins[25], moisture[26,27], polymers[28,29]. Analytical methods have also evolved, including generalized two-dimensional correlation spectroscopy[8], two-trace two-dimensional correlation spectroscopy (2T2D-COS) that extracts useful information from pairs of spectra[15,22,26,30,31], and disrelation mapping applied to hyperspectral imaging in pixel space[28,32,33]. While 2DCOS is valuable for deriving insights from one or two types of spectroscopic data obtained in response to perturbations, its application to three or more different types of multifaceted characterization data has been challenging.

In this study, we propose a general-purpose analytical method suitable for multifaceted characterization data: the tabular two-dimensional correlation analysis. This method is designed to quantitatively evaluate the similarity and sequence of changes using a table composed of multiple structural parameters obtained from various analytical techniques, not limited to spectroscopy. Specifically, the proposed analytical method involves rearranging structural parameters through hierarchical clustering[4,7], followed by calculating and visualizing their phase lags through asynchronous correlations in a heatmap format. This approach allows for a clear visualization of the degree of similarity and sequence among the changes in structural parameters. In this paper, we report on the application of the proposed method to two types of data: typical synthetic data and experimental data extracted from carbon nanotubes (CNTs) that have undergone annealing at high temperatures[34]. Specifically, annealed CNT dataset includes 11 structural parameters measured by 8 different analytical techniques. This illustrates the versatility and effectiveness of the proposed method in handling complex multifaceted characterization datasets derived from a variety of analytical approaches.

**Methods**

Tabular two-dimensional correlation analysis

The proposed tabular two-dimensional correlation analysis comprises four steps: (1) normalization of each structural parameter to a range of 0-1, (2) sorting of structural parameters based on hierarchical clustering using cosine distance as a metric, (3) calculation of asynchronous correlations for each pair of structural parameters, and (4) creation of a dendrogram based on cosine distances and a heatmap where the values of asynchronous correlations are represented by pseudo-colors. Additionally, to validate the computational behavior, we used synthetic data consisting of 8 datasets, each generated by applying a different phase lag of $\pi/4$ to a sine function. This approach served as a test case to confirm the method's efficacy.

Annealed Carbon Nanotube Film Dataset

We utilized a dataset of CNT films annealed at high temperatures, previously reported using a multifaceted analytical approach[34]. In this study, we compared the micro and macro structures using different analytical methods, focusing on a series of datasets comprising 8 methods and 11 structural parameters, described as scalar values (Table 1). This dataset involves films annealed under seven conditions (1200, 1600, 1800, 1900, 2000, 2200, and 2400°C) in a high-temperature furnace under an argon gas atmosphere, evaluated using 8 different measurement methods. The characteristics obtained from each measurement method are listed in the third column of Table 1. The data derived from these measurements encompass a range of scales, including information about defects in the graphite structure, voids between tubes, the structure within and between bundles of multiple tubes, and impurities originating from amorphous carbon.

Table 1. Annealed CNT dataset composed of 11 structural parameters measured by 8 characterization methods.

| Characterization Method | Structural Parameter | Description |
| --- | --- | --- |
| Raman Spectroscopy[35–37] | G/D band ratio | Indicator of carbon defects |
| X-ray Diffraction (XRD)[38–40] | d_intertube(002) | Inter-tube spacing |
| | Lc_interlayer (004) | Crystallite size at inter-layer spacing |
| Wide Angle X-ray Scattering (WAXS)[41,42] | d(10) | (10) hexagonal closed packing of CNT bundles |
| | d(11) | (11) hexagonal closed packing of CNT bundles |
| Far-infrared (FIR) Spectroscopy[43–46] | Effective length | Effective current path length between defects |
| Nitrogen Gas Adsorption[47,48] | Specific surface area | Overall specific surface area |
| | Adsorption amount at low pressure | Adsorption at interstitial sites in inter-tube spacing |
| X-ray Absorption Fine Structure (XAFS)[49–51] | π*/σ* band ratio | $sp^2/sp^3$ carbon ratio |
| Thermo-gravimetric Analysis (TGA)[52] | Weight loss at 500°C | Amount of carbonaceous impurity |
| Positron Annihilation[53,54] | Pore radius | Closed and open nano-spacing |

**Results and Discussion**

The computational flow of the proposed tabular two-dimensional correlation analysis is illustrated in Fig 1. When handling multifaceted characterization data, several unique characteristics differentiate it from typical vibrational spectroscopy data. Firstly, multifaceted characterization data contains structural parameters with vastly different orders of magnitude and units, which, if untreated, could lead to biased data processing. To address this, we normalized each structural parameter to a range of 0 to 1. Next, we tackled the ordering of structural parameters. Unlike traditional vibrational spectroscopy data, structural parameters do not have a predefined sequence like wavelengths, wavenumbers, photon energy, etc. Therefore, we resolved this issue of sequence by sorting structural parameters based on hierarchical clustering, using multidimensional scaling, with cosine distance as one method for calculating similarities between parameters. Finally, we conducted two-dimensional correlation analysis, calculating asynchronous correlations for each pair of structural parameters. The values of asynchronous correlations were used as pseudo-colors to create a heatmap, and the distances between pairs of structural parameters were visualized through a dendrogram. By implementing such a process, it becomes possible to conveniently determine the sequence of changes from asynchronous correlations while keeping high visibility of the similarity of parameters from adjacent positioning, useful for causal inference. This approach facilitates an intuitive understanding of how closely related parameters behave in relation to one another over the change of perturbation.

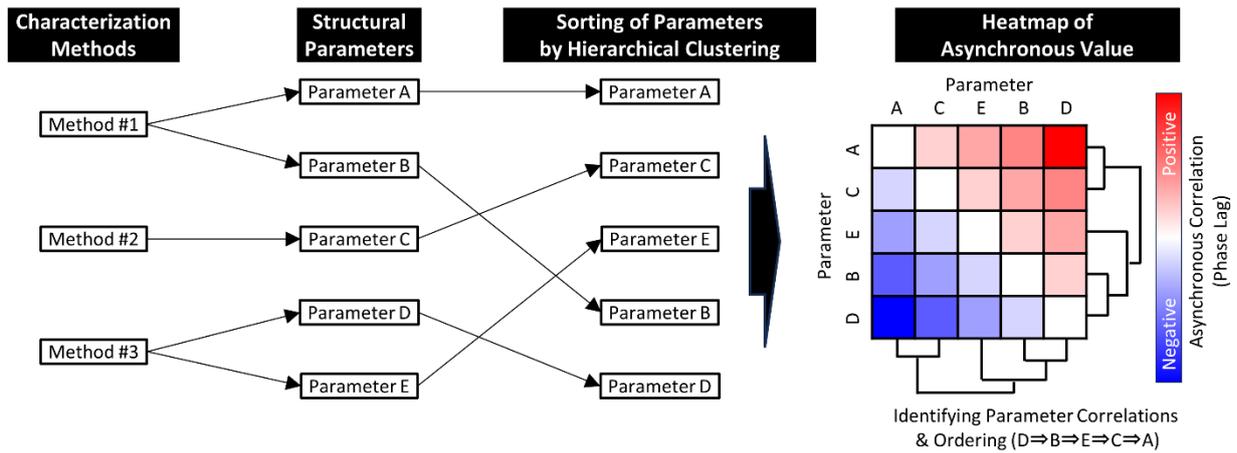

Figure 1. Schematic diagram of the proposed tabular two-dimensional correlation analysis.

The results of calculations performed using the proposed method, applied to a dataset simulating multifaceted characterization data obtained from experiments, are presented in Fig. 2. For validation, we artificially created functions based on the sine function, altering the phase by $\pi/4$ increments, resulting in eight data series labeled #1 to #8, representing perturbations ranging from 0 to $2\pi$ at 16 points (Fig. 2a). We then conducted hierarchical clustering on these data, with Fig. 2b showing a dendrogram representing the cosine distance differences. Adjacent IDs, i.e., synthetic data with small phase differences, are located in close proximity on the dendrogram, forming three major clusters: #1–3, 4–5, and 6–8. Based on these results, we sorted the synthetic data and calculated the asynchronous correlations for each pair, as shown in Fig. 2c. The heatmap reveals clear differences in the phases of the artificially created functions, for example, a significant phase lag between IDs #5 and 8. This demonstrates that our proposed analytical method successfully visualizes the similarity in parameter changes and phase delays.

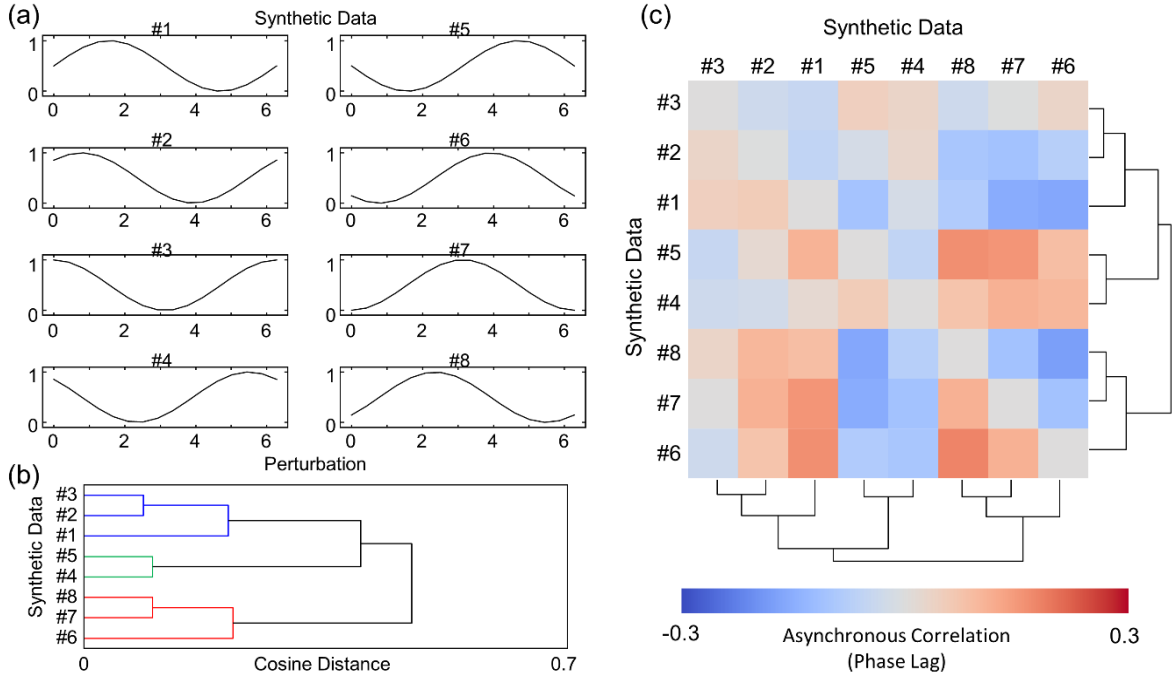

Figure 2. Tabular two-dimensional correlation analysis for synthetic data. (a) 8 synthetic data based on sine function. (b) Dendrogram of 8 synthetic data derived from hierarchical clustering. (c) Heatmap of asynchronous correlation of synthetic data.

Next, we examined the applicability of our proposed method to the carbon nanotube (CNT) data obtained from actual experiments (Fig. 3). CNTs, cylindrical materials made of carbon, possess outstanding mechanical, thermal, and electrical properties. In recent years, their application in transparent conductive films[6,55], supercapacitors[56,57], conductive fibers[5,58–60], and anti-static or reinforcement agents in rubber[61–63] and plastics[64] have been anticipated. However, assemblies of CNTs are highly complex as they are comprised not only of carbon atoms but also a complex combination of hierarchical features as the length-scale increases, such as cylindrical tubes, bundling of multiple tubes, stacking structures, voids within and between tubes and bundles, tortuosity of these voids, carbon impurities, etc. Each of these features influence the properties observed at the macroscopic scale[52]. To understand the structural changes at these different levels, we annealed the CNTs under seven different temperature conditions and evaluated them using 8 analytical methods (Fig. 3a). Through this multifaceted analysis of annealed CNTs, we acquired 11 different structural parameters as outlined in Table 1[34]. Applying the proposed tabular two-dimensional correlation analysis, we investigated the origins of structural changes occurring during high-temperature annealing.

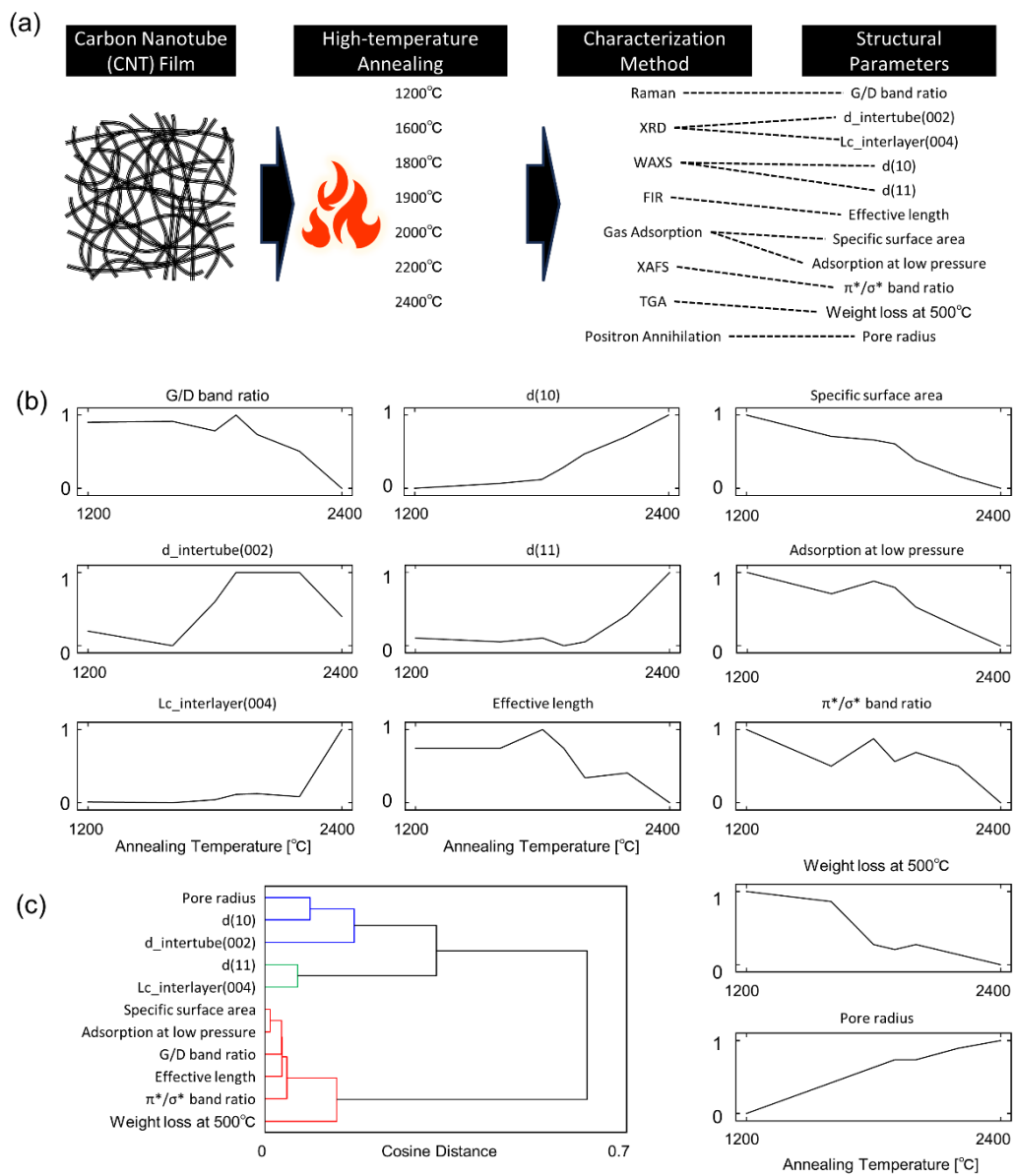

Figure 3. Carbon nanotube film (CNT) dataset with high-temperature annealing. (a) Schematic diagram of annealed CNT film dataset. (b) Effects of the perturbation (annealing temperature) on the 11 structural parameters of annealed CNT films. (c) Dendrogram of the 11 structural parameters in the annealed CNT film dataset, showing the similarity among the structural parameters.

The fluctuations of the 11 structural parameters with annealing temperature as the perturbation are shown in Fig. 3(b). The behavior of parameter changes varies across temperature ranges. For example, void sizes measured by positron annihilation spectroscopy show gradual changes, while parameters corresponding to the (004) interlayer diffraction observed by XRD, indicative of crystal size (Lc_interlayer(004)), undergo abrupt changes in the high-temperature range. These changes, which are complex behaviors, stem from different mechanisms and are difficult to comprehend and/or identify from a simple observation of a graph. To visualize these multiple structural parameter fluctuations, we calculated cosine distances among structural parameters as a parameter similarity measure for hierarchical clustering analysis (Fig. 3(c)). Here, the parameters are broadly divided into three groups. The first cluster includes void size, the (10) plane diffraction related to the packing near the bundle, and the (002) plane diffraction corresponding to the distance between individual CNTs, parameters related to voids. The second cluster contains parameters related to crystal size and the (11) plane diffraction derived from packing at diagonal positions in the bundle, likely corresponding to structural changes accompanying crystal transformations. The third cluster encompasses numerous parameters, including those related to amorphous carbon and some pertaining to microstructure (corresponding to the pore volume) measured by gas adsorption, which are scattered near the cluster related to voids. Although these clusters roughly indicate groups of similar parameters, unraveling which structures change at what stages remains an extremely daunting challenge.

To visualize the pairwise relationships between parameters, we created a heatmap of asynchronous correlations using the proposed tabular two-dimensional correlation analysis in Fig. 4(a). Along with the parameter similarities shown in Fig. 3(c), the visualization of phase lags through asynchronous correlations allows for a discussion on the sequence of parameter changes. For instance, observing the top-right area of the heatmap, the asynchronous correlation between void size and the G/D band ratio takes a positive value, indicating a positive phase lag in the change of void size relative to the G/D band ratio. In other words, the change in void size occurs after the change in the G/D band ratio. By interpreting these asynchronous correlations, we organized the sequence of changes for each structural parameter relative to the annealing temperature (Fig. 4(b)). For example, the changes in parameters derived from thermogravimetric analysis and gas adsorption occurring earlier at lower temperatures suggest that the detachment of amorphous carbon precedes, influencing the $sp^2/sp^3$ carbon ratio and conductive paths. In the subsequent steps, the significant phase lags shown by XRD changes in the (10) and (11) planes suggest alterations in the voids between bundles and crystal size increase, likely due to increases in graphitization. This sequence of structural parameter changes indicates the interplay of two phenomena: the detachment of amorphous carbon and graphitization, thereby estimating the sequence of their impact. Thus, using our proposed method, we successfully extracted valuable information that aids in elucidating the mechanisms impacting microstructures during high-temperature annealing. While this analysis is feasible with a small amount of data, it is not universal but serves as a useful tool for generating insights for those with material knowledge, making it a promising approach in unraveling complex, multifaceted characterization data involving intricate phenomena.

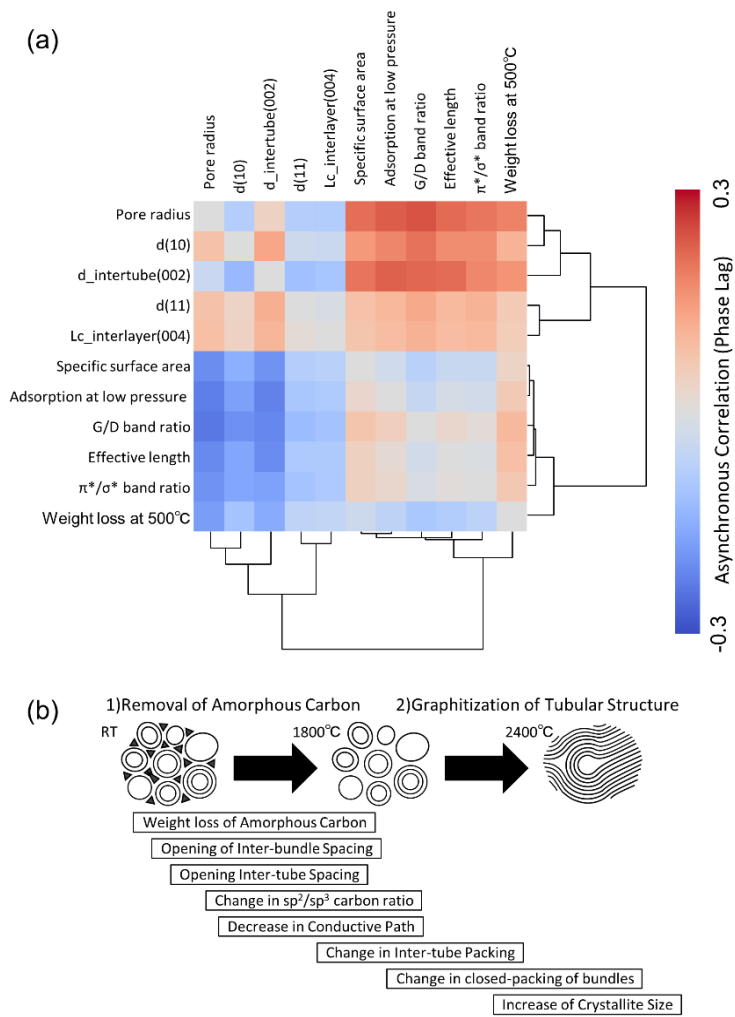

Figure 4. Effectivity and revealed mechanism of annealed CNT dataset based on the proposed method. (a) Heatmap of the asynchronous correlations of the structural parameters in annealed CNT dataset. (b) Mechanism of the high-temperature annealing of CNT derived from the similarities and the phase lags of the structural parameters.

**Conclusion**

We proposed tabular two-dimensional correlation analysis aimed at feature extraction from multifaceted characterization data. This method enables the clear visualization of the similarity in structural parameter changes and phase lags by constructing a heatmap that combines hierarchical clustering with asynchronous correlations of table data comprising different structural parameters. Applying this method to datasets of CNT films annealed at high temperatures, we obtained results that strongly support the understanding of structural origins due to differences in annealing temperatures. The proposed approach is effective even with a small amount of data, applicable to complex data where multiple mechanisms coexist, and aids in uncovering useful information for elucidating mechanisms. We are confident that this technique, while demonstrated on CNTs, can be expanded to a wide range of material analyses.

**Declaration of Conflicting Interests**

The authors declared no potential conflicts of interest with respect to the research.

**Funding**

The authors disclosed receipt of the following financial support for the research: Japan Society for the Promotion of Science (JSPS) KAKENHI (No. JP22K04888, JP22K14571), a project (No. JPNP16010) commissioned by the New Energy and Industrial Technology Development Organization (NEDO), Nanotechnology Platform Program <Molecule and Material Synthesis> (JPMXP09S-16-SH-0027 & S-19-JI-0044) of the Ministry of Education, Culture, Sports, Science and Technology (MEXT).

**Acknowledgments**

We appreciate Dr. Ken Kokubo and Dr. Don N. Futaba for their support. The authors also thank Mr. Hikaru Yorozuya, Mrs. Yukari Urano, Mrs. Junko Kamata for help with experiments. We utilized a large language model for language polishing.

**CRediT authorship contribution statement**

Conceptualization: S.M., Methodology: S.M., Software: S.M., Formal Analysis: S.M., Investigation: S.M., S.Y., K.M., H. N., T.M., N.O., K.K., T.O., Visualization: S.M., K.K. Management: K.K., Supervision: T.O., Writing – Original Draft: S.M.

**ORCID**

Shun Muroga https://orcid.org/0000-0002-6330-0436
Koji Michishio https://orcid.org/0000-0003-1381-7856
Hideaki Nakajima https://orcid.org/0000-0001-8678-0906
Takahiro Morimoto https://orcid.org/0000-0003-0439-7853
Nagayasu Oshima https://orcid.org/0000-0002-5713-112X
Kazufumi Kobashi https://orcid.org/0000-0002-7122-7638
Toshiya Okazaki https://orcid.org/0000-0002-5958-0148